\documentclass[12pt]{article}
\usepackage{amsfonts}
\usepackage{amssymb}
\usepackage{epsfig}
\usepackage{delarray}

\def\simge{\mathrel{%
   \rlap{\raise 0.511ex \hbox{$>$}}{\lower 0.511ex \hbox{$\sim$}}}}
\def\simle{\mathrel{
   \rlap{\raise 0.511ex \hbox{$<$}}{\lower 0.511ex \hbox{$\sim$}}}}

\newcommand{\beq}{\begin{equation}}
\newcommand{\eeq}{\end{equation}}
\newcommand{\al}{\alpha}
\newcommand{\be}{\beta}

\newcommand{\di}{\displaystyle}

\newcommand{\ga}{\gamma}

\newcommand{\scA}{{\scriptscriptstyle A}}

\newcommand{\scU}{{\scriptscriptstyle U}}

\newcommand{\scDM}{{\scriptscriptstyle DM}}

\newcommand{\si}{\sigma}

\renewcommand{\epsilon}{\varepsilon}

\begin{document}

\title{\bf Higgsinos as carriers of SUSY residual effects}
\author{G.M. Vereshkov \footnote{email: gveresh@ip.rsu.ru}, V.A. Beylin, V.I. Kuksa,
R.S. Pasechnik \footnote{email: rpasech@thsun1.jinr.ru}}

\date{Research Institute of Physics,\protect\\
Rostov State University, Rostov-on-Don, Russia}

\maketitle

\begin {abstract}

New version of the MSSM scales is discussed. In this version
$\mu\ll M_{SUSY}\sim M_0\sim M_{1/2}$, where $\mu$ is the Higgsino
mass, $M_0$ is the mass scale of sleptons and squarks, $M_{1/2}$
is the mass scale of gaugino. Renormalization group motivation of
this MSSM version is proposed. The radiation corrections give main
contribution to the splitting of neutralino and chargino masses if
$M_{SUSY}>10^7 GeV$. Calculation of the mass difference has led to
the value $M_{\chi_1^\pm}-M_{\chi_{1,2}^0}\sim 360\, MeV$ at
$M_{\chi}\sim\mu\sim 3\, TeV$. The mean life time of charged
Higgsino in this model is $\tau_{\chi_1^\pm}=0.4\times 10^{-9}\,
s$. The analysis of the residual neutralino concentration in the
cosmological plasma is developed and the value of Higgsino mass is
estimated: $M_{\chi}\approx 3\, TeV$. Formation of the residual
neutralino concentration occurs in the high symmetric phase of the
cosmological plasma. Problems of the relic Higgsinos searches at
the underground laboratory NUSEL and Galactic neutralino
annihilation radiation at the satellite detector GLAST are
discussed. It is shown that the neutralino-nucleon cross section
is spin-dependent. The annihilation of Galactic neutralinos occurs
in $Z^0Z^0$ and $W^{+}W^{-}$ pairs in t- and s-channels, and also
in lepton-antilepton and quark-antiquark pairs in the s-channel.
The contribution to the annihilation spectrum from first two
processes is calculated with using experimental data about hadron
multiplicities at the mass surfaces of $Z^0,W^{\pm}$-bosons; the
calculation of quark-antiquark s-channel contribution is based on
the phenomenological model of hadron multiplicities at
$\sqrt{s}=2M_{\chi}$ with using negative binomial distribution.
\end {abstract}

\section{Introduction.}

As it is known, a theory doesn't fix the MSSM scale hierarchy. Both neutralino
and chargino have different properties depending on this hierarchy, therefore
predictions for experiments are also different. Neutralino is nearly bino in
generally accepted variant of the theory, and neutralino interactions with
matter mainly occur through the scalar quarks' exchange since their masses are
close to the neutralino mass. The splitting between neutralino and chargino
masses is large enough and its value has order a few GeV. It allows to register
these particles through the lepton and hadron cascade decays. The standard MSSM
version implies an existence of the Supergauge Desert between electroweak and
GUT scales. But another variants are also possible. In particular, we are
interested in in-depth study of the MSSM variant, where the mass scales of
squarks and sleptons and the gaugino scale were arranged far from electroweak
scale in multi-TeV region (so-called the Split Supersymmetry) \cite{dimopoulos2},
\cite{dimopoulos1}. Formally Higgsino mass remains arbitrary so it may be
positioned at the EW scale. It's so-called the light Higgsino scenario. We study
in details the properties of this model, which is interesting because in the case
ideas of SUSY are conserved. It should be especially actual if LHC results for
the SUSY effects' observation will be negative.

\section{Objects of investigation and its properties.}

The experimental testing of the supersymmetry hypothesis is one of the general
purposes of Large Hadron Collider \cite{CMS}, \cite{CMS1},
\cite{atlas}. Main object of investigation is the Minimal
Supersymmetric Standard Model, which represents the supersymmetric
generalization of the Standard Model. Experimental program of supersymmetry
searches on the LHC is based on the SUSY breaking scale estimation
$M_{SUSY}\sim 1\;-\;2\, TeV$. There are four theoretical arguments that have lead
to the estimation: 1) the MSSM is the simplest supersymmetric model compatible
with well-known experimental data; 2) the mass spectrum of Higgs bosons is
stable; 3) precise convergence of invariant charges at one point; 4) in the
framework of MSSM we have natural interpretation of Dark Matter.

As it is known, after spontaneous breaking of the gauge symmetry
supersymmetric partners of Higgs and electroweak gauge fields
forms two Dirac electrically charged particles -- charginos
$\chi^\pm_{1,2}$, and four Majorana electrically neutral particles
-- neutralinos $\chi^0_{\al},\;\al=1,2,3,4.$ Lightest
supersymmetric particle $\chi^0_1$ is main candidate to be the
cold Dark Matter component \cite{Primack}, \cite{Primack1},
\cite{ellis}, \cite{jungman}. Generally $\chi^0_1$ is the
superposition of $U(1)$ gaugino $\tilde B$ ("bino"), neutral
$SU(2)$ gaugino $\tilde{W}_{3}$ ("wino") and two Higgsinos $\tilde
h_1^0,\;\tilde h_2^0.$ Structure and properties of chargino and
neutralino depend on relations between characteristic MSSM scales.
Strong theoretical arguments in favor of one or another hierarchy
of scales are absent.

In the next section will be shown that only two variant of MSSM are
theoretically natural. First variant -- the light gaugino scenario (bino or
wino or their mixture), in which \beq \di |\mu|\gg M_{0}\sim M_{1/2}\sim
M_{SUSY}>M_{EW},
\label{gaugino} \eeq -- is well investigated early \cite{pierce},
\cite{gondolo1}, \cite{gondolo2}, \cite{aloisio1}.

We consider second alternative variant of MSSM, in which the Higgsino mass
scale is the nearest for EW scale, but $M_{SUSY}$ is in the multi-TeV region
-- so-called light (separated) Higgsino scenario:
\beq \di M_{0}\sim M_{1/2}\sim M_{SUSY}\gg |\mu| > M_{EW}.
\label{M} \eeq

\section{Renormgroup analysis.}

In our model we shown precise convergence of invariant charges at the scale
compatible with proton mean life. In the known MSSM scheme this convergence of
invariant charges at the $M_{GUT}$ provides the MSSM building-in to the
Supergravity and super string theory. In the case the scale of SUSY breaking is
no so large: $M_{SUSY}\sim M_{1/2}$, so quadratic divergencies cancel near
$M_H$.

We take into account all degrees of freedom in the $SU_{SUSY}(5)$ except of
quark-lepton bosons. It have been found that the states, which are near
$M_{GUT}$, are very important for the renormgroup analysis too.

Initial values of running constants are fixed at the $M_Z$-scale:
\[
\di \al^{-1}(M_Z)=127.922\pm 0.027,\quad \al_s(M_Z)=0.1200\pm0.0028, \quad
\sin^2\theta_W(M_Z)=0.23113\pm0.00015.
\]
To solve renormgroup equations the following initial values are used: \beq
\begin{array}{c}
\di \al_1^{-1}(M_Z)=\frac35\al^{-1}(M_Z)(1-\sin^2\theta_W(M_Z))=59.0132
\mp(0.0384)_{\sin^2\theta_W}\pm(0.0124)_\al,
\\[3mm]
\di \al_2^{-1}(M_Z)=\al^{-1}(M_Z)\sin^2\theta_W(M_Z)=29.5666\pm
(0.0192)_{\sin^2\theta_W}\pm(0.0062)_\al
\\[3mm]
\al_3^{-1}(M_Z)=\al_s^{-1}(M_Z)=8.3333\pm0.1944.
\end{array}
\label{exp} \eeq

In the one-loop approximation invariant charges at the scale $Q_2$ are linked with
its' values at the scale $Q_1$:
\beq \di
\al_i^{-1}(Q_2)=\al_i^{-1}(Q_1)+\frac{b_i}{2\pi}\ln\frac{Q_2}{Q_1},\qquad
b_i=\sum_jb_{ij}. \label{b} \eeq Here all states with masses $M_j<Q_2/2$ at
$Q_2>Q_1$ are summed.

Extra degrees of freedom are taken into account in our analysis, namely:
singlet quarks and their superpartners $(D_L,\,\tilde D_L),\;(D_R,\,\tilde
D_R)$ from superhiggs quintets of $SU_{SUSY}(5)$; gauge superfield
$(\Phi_L,\,\tilde\Phi_L)$ from 24-plet in adjoint representation of $SU(2)$ and
chiral superfield $(\Psi_L,\,\tilde\Psi_L)$ in the adjoint representation of
$SU(3)$. In the minimal $SU_{SUSY}(5)$ we have $M_5=(M_D,\,M_{\tilde D}),\;
M_{24}=(M_{\Psi},\,M_{\tilde\Psi},\,M_{\Phi},\, M_{\tilde \Phi})$ and all of
them are generated by interaction with Higgs condensate at $M_{GUT}$.
Inequality $M_5,\,M_{24}< M_{GUT}$, in principle, can be provided with
precision in $1\;-\;2$ orders.

So, one-loop invariant charges at the scale $q^2=(2M_{GUT})^2$ depend on all
characteristic MSSM scales:
\beq
\begin{array}{c}
\di \al_1^{-1}(2M_{GUT})=\al_1^{-1}(M_Z)-\frac{103}{60\pi}\ln 2+
\frac{1}{2\pi}\left(-7\ln M_{GUT}+\frac{4}{15}\ln
M_{D}+\frac{2}{15}\ln M_{\tilde D}\right.
\\[5mm]
\di \left. +\frac{11}{10}\ln M_{\tilde q} +\frac{9}{10}\ln M_{\tilde
l}+\frac25\ln \mu +\frac{1}{10}\ln M_H +\frac{17}{30}\ln M_t+\frac{53}{15}\ln
M_Z \right),
\\[5mm]
\di \al_2^{-1}(2M_{GUT})=\al_2^{-1}(M_Z)-\frac{7}{4\pi}\ln 2+
\frac{1}{2\pi}\left(-3\ln M_{GUT}+\frac43\ln
M_{\tilde\Phi}+\frac23\ln M_{\Phi}\right.
\\[5mm]
\di \left.+\frac{3}{2}\ln M_{\tilde q} +\frac{1}{2}\ln M_{\tilde l}+\frac43\ln
M_{\tilde W}+\frac23\ln \mu +\frac{1}{6}\ln M_H
+\frac{1}{2}\ln M_t-\frac{11}{3}\ln M_Z \right),
\\[5mm]
\di \al_3^{-1}(2M_{GUT})=\al_3^{-1}(M_Z)+\frac{23}{6\pi}\ln 2+
\frac{1}{2\pi}\left(-\ln M_{GUT}+2\ln M_{\tilde\Psi}+\ln M_\Psi
\right.
\\[5mm]
\di \left.+\frac{2}{3}\ln M_{D}+\frac{1}{3}\ln M_{\tilde D}+ 2\ln M_{\tilde q}
+2\ln M_{\tilde g}+\frac23\ln M_t-\frac{23}{3}\ln M_Z
\right).
\end{array}
\label{gut} \eeq Here $M_0=(M_{\tilde q},\,M_{\tilde l})$ is the "common" mass of scalar quarks
and leptons, averaged over chiralities and generations; $M_t$ is the $t$-quark
mass; others mass parameters were introduced above. In compliance with internal
feature of the MSSM in (\ref{gut}) it supposes that the lightest Higgs boson
mass is close to $M_Z$, but masses of others Higgs bosons $H,\,A,\,H^\pm$ lie
at the $M_H$ scale. Expressions
(\ref{gut}) do not depend on placing of
$M_{24},\,\,M_5,\,M_0,\,M_{1/2},\,\mu,\,M_H$ at the energy scale.
If masses of
singlet superquarks and residual Higgs superfields are equaled to $M_{GUT}$
they are eliminated from renormgroup equations.

At every scale corresponding degrees of freedom are linked and the equaling of
all charges at $M_{GUT}$ leads to equations: \beq
\begin{array}{c}
\di M_{GUT}=Ak_1M_Z\left(\frac{M_Z}{M'_{1/2}}\right)^{2/9},\qquad
\mu=Bk_2M_Z\left(\frac{M_Z}{M'_{1/2}}\right)^{1/3},
\end{array}
\label{3} \eeq where \beq
\begin{array}{c}
\di k_1= K_{\tilde q\tilde l}^{-1/12}K_{GUT1}^{1/3}\equiv
\left(\frac{M_{\tilde l }}{M_{\tilde
q}}\right)^{1/12}\left(\frac{M_{GUT}}{M'_{GUT}}\right)^{1/3},
\\[5mm]
\di k_2=K_{Ht}^{-1/4}K_{\tilde q\tilde l}^{1/4}K_{\tilde g\tilde
W}^{7/2}K_{GUT2}^{-1}\equiv \left(\frac{M_t}{M_H}\right)^{1/4}
\left(\frac{M_{\tilde q }}{M_{\tilde
l}}\right)^{1/4}\left(\frac{M_{\tilde g }}{M_{\tilde
W}}\right)^{7/2}\left(\frac{M''_{GUT}}{M_{GUT}}\right),
\end{array}
\label{4} \eeq \beq
\begin{array}{c}
 \di M'_{1/2}\equiv (M_{\tilde W}M_{\tilde g})^{1/2},\qquad M'_{GUT} \equiv
(M_{\tilde \Psi}M_{\tilde \Phi})^{1/3}(M_{\Psi}M_{\Phi})^{1/6}\leq
M_{GUT},
\\[5mm]
\di M''_{GUT} \equiv \frac{ (M^2_{\tilde \Psi}M_{\Psi})^{7/6}(M_D^2M_{\tilde
D})^{1/2}}{(M^2_{\tilde \Phi}M_{\Phi})^{4/3}}\leq M_{GUT},
\\[5mm]
\di
A=\exp\left(\frac{\pi}{18}(5\al_1^{-1}(M_Z)-3\al_2^{-1}(M_Z)-2\al_3^{-1}(M_Z))-\frac{11}{18}\ln
2\right)=(1.57\times ^{1.09}_{0.92})\cdot 10^{14},
\\[5mm]
\di
B=\exp\left(\frac{\pi}{3}(5\al_1^{-1}(M_Z)-15\al_2^{-1}(M_Z)+7\al_3^{-1}(M_Z))+\frac{157}{12}\ln
2\right)=(2.0\times ^{0.15}_{6.56})\cdot 10^{3}.
\end{array}
\label{5} \eeq

The mass ratios that are labelled as $K$ are quantities, which are
larger than unity. All parameters $K_{GUT1},\,K_{GUT2}$ both in
the MSSM, and in the $SU_{SUSY}(5)$ are not under the theoretical
control. Due to experimental restriction $M_H>315\, GeV$ it is
supposed that $1\le K_{GUT1}\simeq K_{GUT2}\le 10$ and $K_{Ht}$
lies in the region $2\le K_{Ht}\le 10$. Parameters  $K_{\tilde
q\tilde l},\,K_{\tilde g\tilde W}$ are fixed by renormgroup
evolution from $M_{GUT}$ to $M_0,\,M_{1/2}$. We suppose that
$1.5\le K_{\tilde q\tilde l}\simeq K_{\tilde g\tilde W}\le 2.5$.

The studying of $M'_{1/2},\,\mu$ as functions of $M_{GUT}$ is the aim of
analysis:
\beq \di M'_{1/2}(M_{GUT})=(Ak_1)^{9/2}M_Z^{11/2}\times
\frac{1}{M_{GUT}^{9/2}},\qquad \mu(M_{GUT})=
\frac{Bk_2}{(Ak_1)^{3/2}M_Z^{1/2}}\times M_{GUT}^{3/2}. \label{6}
\eeq

We used known restrictions for the proton mean life and for $M_{SUSY}$:
$(\tau_p\geq 10^{32}$ years at $M_{GUT}\geq 10^{15}\, GeV$); $(M_{SUSY}\sim
M'_{1/2}>100\, GeV$ at $M_{GUT}< 3\cdot 10^{16}\, GeV$).

\begin{figure}[!h]
\begin{minipage}{1.3\textwidth}
\epsfxsize=\textwidth\epsfbox{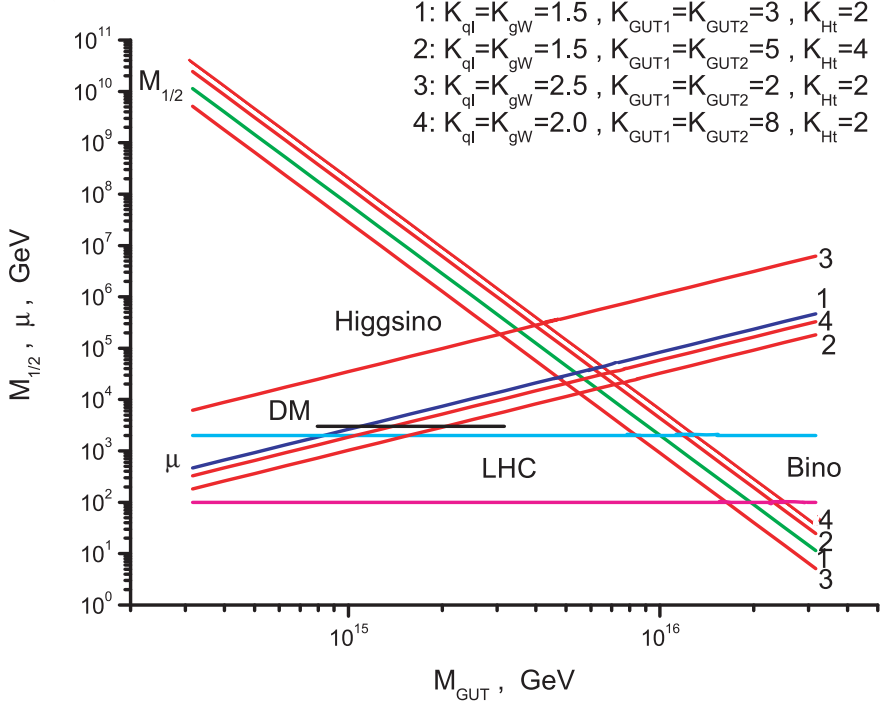}
\end{minipage}
\caption{\small Two variants of the MSSM scales -- bino-like LSP $\mu \gg M'_{1/2}$
and Higgsino-like LSP $\mu \ll M'_{1/2}$.}
\label{mod}
\end{figure}

From the analysis two variants of parameters (\ref{6}) were found. In the first case $\mu
\ll M'_{1/2}$, in the second one $\mu \gg M'_{1/2}$ that is shown at Fig. \ref{mod}.

\section{Analysis of mass spectrum.}

Formally the $M_H$ scale may be arbitrary. In this scenario light
particles $\chi^0_{1,\,2}$ and $\chi^\pm_1$ have a structure of Higgsino with
masses \beq
\begin{array}{c}
\di M_{\chi^0_1}\simeq |\mu|-\frac{M_Z^2(1+sign(\mu)\sin
2\be)}{2}\left(\frac{\cos^2\theta}{M_{\tilde W}}+\frac{\sin^2\theta}{M_{\tilde
B}}\right)\approx |\mu|-\frac{M_{Z}^2}{2 M_{SUSY}}
\\[5mm]
\di |M_{\chi^0_2}|\simeq |\mu|+\frac{M_Z^2(1-sign(\mu)\sin
2\be)}{2}\left(\frac{\cos^2\theta}{M_{\tilde W}}+\frac{\sin^2\theta}{M_{\tilde
B}}\right)\approx |\mu|+\frac{M_{Z}^2}{2 M_{SUSY}}
\\[5mm]
\di M_{\chi^\pm_1}\simeq |\mu| - \frac{M_W^2}{M_{\tilde
W}}\left(\frac{|\mu|}{M_{\tilde W}}+ sign(\mu)\sin 2\be
\right)\approx |\mu|.
\end{array}
\label{H} \eeq Heavy particles $\chi^0_{3,\,4}$ and $\chi^\pm_2$
have a structure of gaugino, their masses lies at $M_{SUSY}$ scale:
\beq \di M_{\chi^0_3}\approx M_{\tilde B},\qquad
M_{\chi^0_4}\approx M_{\tilde W} ,\qquad M_{\chi^\pm_2}\approx M_{\tilde W}.
\label{G} \eeq

Interactions through virtual gaugino, squarks and sleptons and heavy Higgs are
suppressed by large masses in denominators. Basic physical effects describing
by the Lagrangian of light Higgsino $\chi^0_{1,\,2}\,,\; \chi^\pm_1$
interactions with photons and vector bosons: \beq
\begin{array}{c}
\di \Delta
L=\left(eA_\mu-\frac{g_2}{2\cos\theta}(1-2\sin^2\theta)Z_\mu
\right) \bar \chi^-_1\ga^\mu\chi^-_1 +
\frac{g_2}{2\cos\theta}Z_\mu(\bar \chi^0_{1}\ga^\mu \chi^0_{2}
+\bar \chi^0_{2}\ga^\mu \chi^0_{1})+
\\[5mm]
\di +\frac{g_2}{\sqrt{2}}W^+_\mu(\bar \chi^0_1+\bar
\chi^0_2)\ga^\mu \chi^-_1+\frac{g_2}{\sqrt{2}}W^-_\mu \bar
\chi^-_1 \ga^\mu (\chi^0_1+ \chi^0_2).
\end{array}
\label{LH} \eeq

Tree constraint on the mass splitting between chargino and
neutralino $M_{\chi^\pm_1}-M_{\chi^0_1}\gtrsim m_e $ with
(\ref{H}) gives some constraint on supersymmetry breaking scale:
$M_{SUSY}< 10^7\, GeV$. But if $M_{SUSY}> 10^7\, GeV$ the
radiative corrections dominate in formation of the
neutralino-chargino mass spectrum. Combination of all mass
operators in the one-loop approximation may lead to the finite
integral on the mass surface:
\[\Delta
M_\chi=-\frac{ie^2M_{Z}^2}{8\pi^4}\int\frac{(\hat{q}-M_{\chi})dq}
{q^2(q^2-M_{Z}^2)[(q+p)^2-M_{\chi}^2]}\;,
\] For $M_{\chi}\gg M_Z$ we have
\beq \Delta M_\chi \simeq \frac{\alpha(M_{\chi}) M_Z}{2}
\label{delta} \eeq To obtain the value $\alpha$ at $M_{\chi}$
scale by renormgroup methods we use estimation $M_{H}\sim M_t\sim
200\, GeV$ \cite{Hboson}, \cite{belyaev} and $M_{\chi}\sim 3\,
TeV$ that will be obtained below. We found $\Delta M_\chi \simeq
360\, MeV$. The calculation of chargino mean life time gives \beq
\di \tau_{\chi_1^\pm}= \frac{15\pi^3}{G_F^2}(\Delta
M_\chi)^{-5}=0.4\times 10^{-9}\;s\,. \label{t} \eeq

If we refuse from Supergauge Desert we haven't theoretical
considerations on the separated Higgsino mass. Nevertheless we can
attempt to make an important step to solve this problem when we
use the hypothesis that neutralino is a carrier of Dark Matter in
the Universe.

\section{Cosmological estimation of the separated neutralino mass.}

In discussing versus of the theory physics of neutral and charged
Higgsinos fixes by the Lagrangian (\ref{LH}) and the mass spectrum
(\ref{H}) with the radiation splitting (\ref{delta}). Therefore any
quantitative predictions of the theory depend on two parameters
$\mu,\;M_H$ only. Concrete value $M_H$ isn't essential. The basic
parameter of the model with separated Higgsino is $\mu.$ Let imply
that neutralino is a carrier of Dark Matter in the Universe.
Irreversible neutralino annihilation starts in the cosmological
plasma at the moment $t_0$ and the temperature $T_0$, when mean energy
of relativistic quarks and leptons compares with neutralino mass:
$\bar \varepsilon_f \simeq 3\,T_0=M_{\chi}.$ Residual density of
neutralino at $t\gg t_0$ describes by the asymptotic solution of
evolutional equations: \cite{Primack}, \cite{Primack1}, \cite{gondolo}:
\beq
\begin{array}{c}
\di \frac{dn_{\chi}}{d t}+3Hn_{\chi}=-\frac12 n_{\chi}^2(\si
v)_{ann},
\\[5mm]
\di 3H^2=8\pi GwT^4, \qquad H=\frac1a \cdot \frac{da}{dt}, \quad
wT^4=\frac{const}{a^4}.
\end{array}
\label{nchi} \eeq Here $(\si v)_{ann}$ -- kinetic cross section of the
annihilation, which approximate to the constant in our model; $G$
-- gravitational constant; $w=w(T)$ -- statistical weight of
plasma at the annihilation epoch. With all degrees of freedom in
the two-doublet Standard Model we have $w=443\pi^2/120$. This
value was used us for analysis of annihilation at the temperature
exceeding the characteristic scale of Standard Model.

The asymptotic solution of equations (\ref{nchi}) at times,
exceeding the moment when annihilation terminates, is \beq \di
n_{\chi}(t)=\left(\frac{a(t_0)}{a(t)}\right)^3\left(\frac{2\pi
Gw}{3}\right)^{1/2} \frac{4T_0^2}{(\si v)_{ann}} \label{nchi1}
\eeq where $T_{0}\simeq M_\chi/3$ is the temperature at the moment of
annihilation beginning. Annihilation epoch begins at the moment
\beq \di t_{0}\simeq \frac{1}{4T_{0}^2}\left(\frac{3}{2\pi
Gw}\right)^{1/2} \label{tann} \eeq and ends when $T_{1}\simeq
M_\chi/20$. To obtain the value of mass density of the neutralino
gas in the present-day Universe $\rho_{\chi}=M_{\chi}n_{\chi}$ we
use the standard data for cosmological neutrino. Evolution of the
neutrino component in the cosmological plasma may be considered as
adiabatic since the energy isolation occurs mainly in non-neutrino channels of the
hadronization and annihilation processes at $t>t_0$. Therefore the
ratio of scale factors in (\ref{nchi1}) we can replace by ratio of
neutrino gas temperatures: \[ \di
\frac{a(t_0)}{a(t_{\scU})}=\frac{T_{\nu}}{T_0}, \qquad
T_{\nu}\simeq \left(\frac{4}{11}\right)^{1/3}T_{\ga}= 1.676\times
10^{-13} \, GeV,
\]
Here $t_{\scU}$ is the age of the Universe; $T_{\nu}\equiv
T_{\nu}(t_{\scU})$, $T_{\ga}$ is the temperature of the relic gamma
radiation.

Finally for density of stable neutralinos in the epoch,
characterizing by the temperature of relic neutrino $T_\nu$, we
obtain: \beq \di
\rho_\chi(T_\nu)=\frac{M_\chi}{T_{0}}\left(\frac{2\pi
Gw}{3}\right)^{1/2} \frac{4T_\nu^3}{(\si v)_{ann}}, \label{DM}
\eeq

From the recent WMAP data \cite{verde}, \cite{spergel} the mass
density of Dark Matter in the present-day Universe is equal: \beq
\di \rho_{\scDM}(t_{\scU})\simeq (0.23\pm 0.04) \rho_c \simeq
(0.94\pm 0.34)\times 10^{-47}\; GeV^4. \label{DMexp} \eeq Here $\di
\rho_c=(4.1\pm0.8)\times 10^{-47} \; GeV^4$ is the critical density
of the Universe. An assumption about the Dark Matter in the
Universe is the neutralino gas provides in \beq
\rho_{\chi}(t_{\scU})=\rho_{\scDM}(t_{\scU}).\label{ysl}\eeq

To estimate the value of neutralino mass $M_{\chi}$ from
(\ref{ysl}) we should obtain the expression for kinetic cross
section of neutralino annihilation in the cosmological plasma
$(\si v)_{ann}$ as a function of $M_{\chi}.$ The problem is that
we don't know in which phase of the cosmological plasma neutralino
annihilation occurs. We analyze two scenario consistently.

{\bf 1. Annihilation in the low symmetric phase.} Formation of the
residual neutralino concentration occurs in the low symmetric phase
(LS-phase) of the cosmological plasma that is after the electroweak
transition if neutralino mass $M_\chi<3T_{eW}\sim 300 \;GeV$. In
standard scenarios of the cosmological evolution of neutralino as
gaugino (\ref{gaugino}) in analysis of the Majorana neutralino
annihilation to massless fermions for execution of the Pauli's
principle we must take into account of thermal corrections as a
result the full annihilation cross section is a function of the
temperature. Formally we have two variants: the annihilation of
short-living particles ($\tau_\chi<t_0$) and the coannihilation of
long-living particles ($\tau_\chi\gg t_0$) \cite{gondolo2}. The
carried out analysis shown that for existing of the separated Higgsino
scenario we must forbid the s-channel annihilation into quarks and
leptons that is to make second Majorana neutralino $\chi_2^0$
unstable with mean life time is smaller than $10^{-10}\;s$. It is possible if
$M_{SUSY}$ scale doesn't exceed $10^{5} \;GeV$. But this scenario
assumes the fine tuning of the neutralino mass to W-boson mass.
Therefore
\[
M_{\chi_1^-}\sim M_{\chi_1^0}\sim M_W,
\]
that is contrary to well known experimental constraints
\cite{hagiwara}. Formally equations for annihilation in the LS-phase
contain the second solution corresponding to neutralino
mass in order a few TeV. But such scale is incompatible with the
assumption about neutralino annihilation in the LS-phase
of the cosmological plasma. So this assumption doesn't reduce to
self-consistent result.

{\bf 2. Annihilation in the high symmetric phase.} As it is known, at
temperatures, exceeding the temperature of electroweak transition
$T>T_{eW}\sim 100\;GeV$, plasma is in the high symmetric phase (HS-phase),
in which Higgs condensate is absent. The characteristic feature of
the HS-phase is the massless of all particles (more precisely,
their masses is $m\ll T$) except Higgsino. Physical states here is
quants of gauge fields $B,W_a \,(a=1,2,3)$ and chiral fermions.
The mass splitting disappear in the Higgsino family. As a result
neutralino and chargino degrees of freedom incorporate to the
fundamental representation of the $SU(2)$ group that is the unified Dirac
field $\chi$, in which internal states classify over quantum
numbers of the restored $SU(2)$ symmetry and they are dynamic
equivalent. At the HS-phase instead of (\ref{LH}) we should use the
Lagrangian \beq \di \Delta L_{\chi}=
\frac{1}{2}g_1B_\mu\bar\chi\ga^\mu\chi+\frac{1}{2}g_2W^a_\mu\bar\chi\ga^\mu\tau_a\chi,
\label{LH1} \eeq which adds to the Standard Model Lagrangian
written in terms of gauge and chiral fields: \beq
\begin{array}{c}
\di L_{SM}= -\frac{1}{2}g_1B_\mu\bar l_L\ga^\mu l_L-g_1B_\mu\bar
e_R\ga^\mu e_R+\frac{1}{6}g_1B_\mu\bar q_L\ga^\mu
q_L+\frac{2}{3}g_1B_\mu\bar u_R\ga^\mu u_R-\frac{1}{3}g_1B_\mu\bar
d_R\ga^\mu d_R
\\[5mm]\di
+\frac{1}{2}g_2W^a_\mu\bar l_L\ga^\mu\tau_a l_L
+\frac{1}{2}g_2W^a_\mu\bar q_L\ga^\mu\tau_a q_L.
\end{array}
\label{LSM1} \eeq

Irreversible Higgsino annihilation occurs in the HS-phase and
governed by the Lagrangians (\ref{LH1}) and (\ref{LSM1}) if
Higgsino mass is $|\mu|\gg T_{eW}$. We calculated all cross
sections of the t- and s-channel Higgsino annihilation into gauge
bosons and massless fermions. The technology of these calculations
is analogous to calculations in the QCD. The difference is we must
take into account the annihilation channels, in which initial and
final states are arbitrary over the two-dimensional color,
corresponding to broken $SU(2)$. Full list of all annihilation
channels is \beq
\begin{array}{c}
\di \chi\chi \to BB,\qquad \chi\chi \to W_aW_a ;
\\[5mm]
\di \chi\chi \to B^* \to l_L\bar l_L,\;e_R\bar e_R,\;q_L\bar
q_L,\;u_R\bar u_R,\;d_R\bar d_R; \qquad \chi\chi \to W_a^* \to
l_L\bar l_L,\;q_L\bar q_L,
\end{array}
\label{HS} \eeq
where $l_L,\;q_L,\;e_R,\;u_R,\;d_R$ is chiral fermions and quarks of three
generations. The calculation of the total kinetic cross section over all
channels (\ref{HS}) gives: \beq \di (\si v)_{ann}=
\frac{21g_1^4+6g_1^2g_2^2+39g_{2}^4}{512\; \pi M_{\chi}^2} \label{sHS} \eeq
Here $g_1=g_1(2M_{\chi}),\,g_2=g_2(2M_{\chi})$ -- gauge coupling constants at
the scale $\sqrt{s}=2M_{\chi}$. Their may be calculated by methods of
renormalization group analysis.

From comparison of the theoretical expression for density
(\ref{DM}) in the assumption, that formation of the residual
neutralino concentration occurs in the HS-phase, with well known experimental
constraints (\ref{DMexp}) follows that separated neutralino can
constitute the Dark Matter in the Universe if its mass \beq
M_\chi=2900\pm 500 \;GeV. \label{Mchi} \eeq With such mass value
the irreversible annihilation starts at the temperature $T_0\sim
M_\chi/3 \sim 1000\;GeV$, and finishes at the temperature $T_1\sim
M_\chi/20\sim 100\;GeV$. Hence the whole process of the
annihilation occurs in the HS-phase that demonstrates
the internal consistency of the neutralino mass evaluation.

Of course, we should treat carefully to the value (\ref{Mchi}) since
nobody knows what the Dark Matter is in reality. Besides this
value supposes the execution of $M_\chi\ll M_{SUSY}$. However, we
should emphasize that the discussing variant of the theory no less
motivated phenomenologically than well known scenarios with
$M_{SUSY}\sim 1\;-\;2 \;TeV$.

\section{Problem of separated neutralino searches.}

Above variant with separated Higgsinos can't be tested at LHC because their mass
splitting is drastically small. So we cannot see any lepton decays of chargino
and neutralino. A check-up of our model is possible now only in astrophysics.

\subsection{Direct detection.}

\subsubsection{Elastic scattering of separated neutralino on nucleons and nuclei.}

There are approximately twenty experimental program for relic WIMPs direct
detection \cite{search1}, \cite{smith1}, \cite{smith2}, \cite{sumner}. But we
haven't any clear evidence of their existence now.

All operating detectors use the registration of nuclear recoils. They are generated
in the processes of the elastic WIMP scattering by nuclei \cite{WIMPSearch}.
As usual, the main problem is to discriminate signal and
background. Now there is a project to use cryogenic apparatus with mass near
$1$ ton in the NUSEL \cite{NUSEL1}, \cite{NUSEL2}, \cite{NUSEL3}, where the muon background
is negligibly small. These experiments are planned in the same period that the LHC.

The most perspective technology is using the liquid xenon as scintillator which react to
passing nuclear recoils. To study the possibility of neutralino detection in such
experiments we consider neutralino-nucleon elastic scattering. It is known
that spin independent part of the total cross section induced by neutralino-quark
interaction through scalar quark exchange \cite{WIMPSearch}. In our model
this part of cross section is strongly damped by large scalar quark masses. As
a result in the cross section we have only spin dependent part in the form: \beq \si_{\chi
n}=\frac{g_2^4 m_n^2}{64\pi m_W^4}\,,\qquad \si_{\chi p}=\frac{g_2^4
(1-4\sin^2\theta_W)^2 m_p^2}{64\pi m_W^4}. \label{SD}
\eeq

As it is known the local DM density is $0.3 \;GeV/sm^3,$ the velocity
of neutralino at the Sun position is near $200 \;km/s.$ Using these data we estimated
the separated neutralino flux in detector $j_{\chi}\simeq 2\cdot 10^3\, \;sm^{-2}\;s^{-1}.$
As example we considered one of the most perspective detectors XENON
which is in the development stage. The detector threshold is $\lesssim 10\;KeV$
for nuclear recoil energy. The typical recoil energy from interaction with galactic separated
neutralino is $\sim 100\;KeV$. Detector contains  $\sim 0.34\;m^3$ of liquid xenon $A=131.3.$
We calculated the elastic cross section of nonrelalivistic neutralino on nuclei of xenon
$\sim 10^{-35}\;sm^2.$ The estimation of number of events gives the value that
close to the planned background for considered detector $4.5\cdot 10^{-5} \;s^{-1}$ or
$\sim 1$ event per hour that can't provide the reliable signal for existing of Dark Matter
in the form of separated neutralino. The question about direct detection possibility of
separated neutralino remains open.

\subsubsection {Recharging of separated neutralino on nucleons.}

New and exotic processes could be possible when relic neutralino penetrate through the
matter. Corresponding channels are:
\[\chi^0+p\to \chi^++n\,,\qquad \chi^0+n\to \chi^-+p\,.\]
Cross sections have the form:
\[\begin{array}{l} \di \si^{*}_{\chi p}=\frac{g_2^4 m_p^2}{16\pi
m_W^4}\sqrt{\frac{E_N-\Delta m_N-\Delta M_\chi}{E_N}},
\\[5mm] \di \si^{*}_{\chi
n}=\frac{g_2^4 m_p^2}{16\pi m_W^4}\sqrt{\frac{E_N+\Delta
m_N-\Delta M_\chi}{E_N}}\,.\end{array}\] Here $\Delta
m_N=m_n-m_p\simeq 1 \, MeV$ -- proton-neutron mass difference;
$\Delta M_\chi=M_{\chi_1^{\pm}}-M_{\chi^{0}} \simeq 360\; MeV$ --
chargino-neutralino mass difference that can be calculated from
(\ref{delta}); $E_N = \frac{1}{2} m_N v_{rel}^2$. An average
kinetic energy of neutralino in the locality of the Sun is about
$\sim 1\;MeV$, but recharging reaction is possible only if there
are high energy neutralinos with $E_{kin}>1\;TeV$ in cosmic rays.

\subsection{Indirect detection.}

\subsubsection{Diffuse gamma ray spectrum.}

Except of direct relic neutralino detection experiments there are many projects use
satellite experiments for the neutralino annihilation products observation
\cite{search1}, \cite{sumner}, \cite{ann}. It can be possible to establish some
qualitative and quantitative WIMP's characteristics from these data, mainly
from the data on the neutralino annihilation spectrum from the Galactic halo. We
consider this process and calculate characteristic energies in the diffuse
gamma spectrum. Dirac neutralinos in the halo annihilate into W- and Z-bosons
\beq \di \chi\chi \to W^{+}W^{-},\qquad \chi\chi \to ZZ,
\label{wz} \eeq and into fermions \beq \chi\chi \to q\bar q,\; l\bar l.
\label{ff} \eeq Corresponding total kinetic cross section given in the form:

\beq \di (\si v)_{ann}=
\frac{g_2^4\,(21-40\cos^2{\theta_W}+34\cos^4{\theta_W})}{256\; \pi
M_{\chi}^2\cos^4{\theta_W}} \label{annhalo} \eeq

It is known that total multiplicity of secondary hadrons nearly twice larger
than charged hadrons multiplicity \cite{hagiwara}, \cite{L3}, \cite{dremin}.
Average multiplicity of charged secondary hadrons $\langle
n_{ch}\rangle(\sqrt{s})$ was studied in $e^+e^-$, $p \bar p$ and $p p$
reactions, $e^{\pm} p$ interactions \cite{hera}. It was established that
charged multiplicity is an universal function of energy
\[ \di \tilde n_{ch}(\sqrt{s})=A+B\ln \sqrt s+C\ln^2\sqrt s,\quad\tilde
n_{ch}\equiv \langle n_{ch}\rangle (\sqrt{s}/q_0)-n_0 \] with fixed parameters
\[ A=3.11\pm 0.08,\quad B=-0.49\pm 0.09,\quad C=0.98\pm 0.02.
\] To choice a specific channel it is need to fix parameters $q_0,\,n_0$.
For neutralino annihilation $\di q_{0(\chi\chi)}=1,\; n_{0(\chi\chi)}=0.$

We have used experimental data on multiplicity in $Z \to hh$ decay
\cite{review}. For estimation we suppose that charged hadrons part $\varkappa
\equiv \langle n_{ch}\rangle /\langle n_{h} \rangle$ doesn't depend on energy
and have the value $\varkappa\simeq 0.49.$

To plan an experiment it is important to know characteristic photon energies
that are generated in the following decays: $\pi^0 \to 2\ga ,\;\eta^0 \to 3\ga
,\;\eta^0\to 3\pi^0 \to 6\ga$. In the annihilation channel
(\ref{ff}) the total hadron multiplicity is described by logarithmic function well:
\beq \di \langle n^{ff}_h \rangle=\varkappa^{-1}(A+B\ln 2M_\chi+C\ln^22M_\chi)\simeq
149\,,\qquad M_\chi\simeq 3 \;TeV. \label{nff}\eeq Average energy of the neutral pion
in neutralino annihilation products is $\bar E_{\pi^0}\simeq \bar
E_{\eta^0}\simeq 2M_{\chi}/\langle n^{ff}_h \rangle$. So we have characteristic
maximal photon energy in the  $\pi^0\to 2\ga$ decay: \beq \bar E_{\ga
(\pi^0\to\, 2\ga)}\simeq \bar E_{\pi^0}/2=20 \;GeV.\label{xe1}\eeq

Analogously from decays $\eta^0\to 3\ga$ and $\eta^0\to 3\pi^0\to 6\ga$ energies
of photons are: \beq \bar E_{\ga
(\eta^0\to\, 3\ga)}\simeq \bar E_{\eta^0}/3=13.5 \;GeV\,,\qquad \bar E_{\ga
(\eta^0\to\, 6\ga)}\simeq \bar E_{\eta^0}/6=6.7 \;GeV. \label{xe2} \eeq

Then we considered second annihilation channel (\ref{wz}). Total hadron
multiplicity of W- and Z-bosons decays is \beq \di
\langle n^{WZ}_h \rangle\simeq 42.9.\label{nwz} \eeq
Average energy of the neutral pion is $\bar E_{\pi^0}\simeq \bar
E_{\eta^0}\simeq M_{\chi}/\langle n^{WZ}_h \rangle$.
So we have characteristic maximal photon energies:
\beq
\begin{array}{c} \bar E_{\ga (\pi^0\to\, 2\ga)}\simeq \bar E_{\pi^0}/2\simeq 35
\;GeV;
\\[5mm] \label{xe3}
\bar E_{\ga (\eta^0\to\, 3\ga)}\simeq \bar E_{\eta^0}/3=23.3
\;GeV \,,\quad \bar E_{\ga (\eta^0\to\, 6\ga)}\simeq \bar
E_{\eta^0}/6=12 \;GeV. \end{array} \eeq

Multiplicity distribution in a wide energy region is described by the negative
binomial distribution (NBD) with a good precision, which depends on energy very weak
(as logarithm) \cite{L3}: \beq
\begin{array}{c}
\di P(n; \tilde n, k)=\frac{k(k+1)...(k+n-1)}{n!}\cdot
\frac{(\tilde n/k)^n}{[1+(\tilde n/k)]^{n+k}};
\\[5mm]
\di k^{-1}(\sqrt{s})= a+b\ln \sqrt s,
\end{array}
\label{nbd} \eeq where $n\equiv n_{ch},\,\tilde n\equiv \tilde
n_{ch}$. Coefficients of the function $k^{-1}(\sqrt{s})$ is different for various channels:
\beq \di a_{e^+e^-}= -0.064\pm 0.003,\qquad
b_{e^+e^-}=0.023\pm 0.002;\label{abee}\eeq \beq \di a_{pp/\bar
pp}= -0.104\pm 0.004,\qquad b_{pp/\bar pp}=0.058\pm
0.001.\label{abpp} \eeq

\begin{figure}[!h]
\begin{minipage}{0.8\textwidth}
\epsfxsize=\textwidth\epsfbox{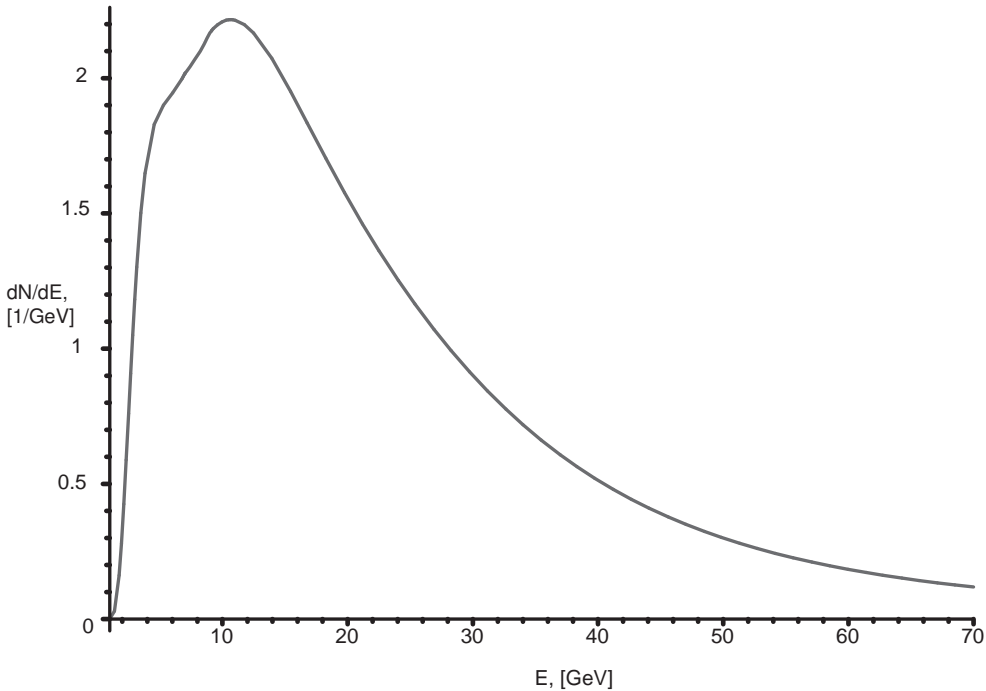}
\end{minipage}
\caption{\small Diffuse gamma ray annihilation spectrum.}
\label{grdN/dE}
\end{figure}

Using NBD we found approximate photon distribution over the energy.
The number of photons with the energy $E_{\ga}$ in one
$\chi\bar\chi$-annihilation act may be defined by \beq
\begin{array}{c}
\di \frac{dN_\ga}{dE_\ga}\simeq \frac{2M_\chi}{E_\ga^2}\left\{\di
\left[Br(\pi^0/h)+ Br(\eta^0/h)\cdot Br(\eta^0\to
2\ga)\right]\times \right.
\\[5mm]
\di \times \left[Br(h)\langle n^{ff}_{ch}\rangle
P\left(\frac{M_\chi}{2E_\ga}; \langle n^{ff}_{ch}\rangle,
k_{ff}\right)+\frac12 Br(WZ)\langle n^{WZ}_{ch}\rangle
P\left(\frac{M_\chi}{4E_\ga}; \langle n^{WZ}_{ch}\rangle,
k_{WZ}\right)\right]+
\\[5mm]
\di+ Br(\eta^0/h)\left[Br(\eta^0\to 3\pi^0)+\frac13 Br(\eta^0\to
\pi^+\pi^-\pi^0)\right]\times
\\[5mm]
\times \di \left[Br(h)\langle n^{ff}_{ch}\rangle
P\left(\frac{M_\chi}{6E_\ga}; \langle n^{ff}_{ch}\rangle,
k_{ff}\right)+\frac12 Br(WZ)\langle n^{WZ}_{ch}\rangle
P\left(\frac{M_\chi}{12E_\ga}; \langle n^{WZ}_{ch}\rangle,
k_{WZ}\right)\right]+
\\[5mm]
\di +\frac13 Br(\eta^0/h)Br(\eta^0\to \pi^+\pi^-\ga)\times
\\[5mm]
\di \left.\times \left[Br(h)\langle n^{ff}_{ch}\rangle
P\left(\frac{M_\chi}{3E_\ga}; \langle n^{ff}_{ch}\rangle,
k_{ff}\right)+\frac12 Br(WZ)\langle n^{WZ}_{ch}\rangle
P\left(\frac{M_\chi}{6E_\ga}; \langle n^{WZ}_{ch}\rangle,
k_{WZ}\right)\right]\right\}. \label{dN/dE}
\end{array}
\eeq Here $Br(WZ)\simeq 0.2$ is the total branching for neutralino annihilation
into $W$- è $Z$-bosons; charge hadron multiplicities $\langle n^{ff}_{ch}\rangle$ and $\langle
n^{WZ}_{ch}\rangle$ defined in (\ref{nff}) and (\ref{nwz});
values $k^{-1}_{ff}=k^{-1}(2M_{\chi})=0.4$ and
$k^{-1}_{WZ}=k^{-1}(M_{\chi})=0.12$ defined with coefficients
(\ref{abpp}) and (\ref{abee}) correspondingly. The spectrum is shown graphically at Fig. \ref{grdN/dE}.

Calculated gamma-ray flux of annihilation radiation from clumped galactic halo
is closes to sensitivity threshold of the satellite detector GLAST. The question about
principal possibility for registration of the neutralino annihilation spectrum at GLAST
remains open.

\subsubsection{High energy gamma ray bursts as a result of collapse
and total annihilation of neutralino clumps.}

Consideration neutralino as a carrier of DM mass in the Universe generates
questions about their evolution and about astrophysical phenomenons which
evidence both the existence of neutralino and the reality of the evolution process
with neutralino participation.

According to high resolution cold dark matter simulations in
framework of hierarchical principle of structure formation large
virilized halos are formed through the constant merging of smaller
halos formed at earlier times. Neutralinos at the gravitational
isolation stage $z=8 \div 10$ formed dense massive clumps of
different scales \cite{small-scale}, \cite{CDM}. It likes
naturally that DM clumps are in the state of dynamical equilibrium
as a result of compensation of own gravitational forces by forces
of tidal interactions with each other. In consequence of inner
instability the clump can turn into irreversible collapse state.
The growth of density provides the growth of annihilation rate. So
neutralino clouds become more and more intense sources of
relativistic particles. At final stage of this process there is a
gamma ray burst with maximal characteristic photon energies
defined above (\ref{xe1}), (\ref{xe2}), (\ref{xe3}). Specific
energy and time distribution of the gamma radiation are connected
with neutralino parameters and give an important information about
structure of the MSSM and about its correspondence with cosmology.
Therefore our main statement consist in the nature of rare
powerful cosmological gamma ray bursts is a result of the total
annihilation of neutralino clumps.

We analyzed the qualitative and quantitative descriptions of this
process in the spherically symmetric collapse model with two
assumptions. First, annihilation products leave a clump during a
time substantially smaller than the time of its macroscopic
evolution. And second, an annihilating clump is spatially
homogeneous and isotropic since more dense regions annihilate
faster than the rest and heterogeneities are vanishing.

The model, defined by equations of mass and momentum balance and the equation of
nonrelativistic gravitational theory, is:
\beq
\begin{array}{c}
\di \frac{d}{dt}\int \rho_{\chi} dV = -\int \rho_{\chi} v_i dS_i -
\int \frac{2(\si v)_{ann} \rho_{\chi}^2}{M_{\chi}}dV\,,
\\[5mm]
\di \frac{d}{dt}\int \rho_{\chi} v_i dV = -\int (\rho_{\chi} v_i
v_k + P_{ik})dS_k - \int \rho_{\chi} \nabla_i \phi dV - \int
\nu_{dis}\rho_{\chi} v_i dV - \int \frac{2(\si v)_{ann}
\rho_{\chi}^2}{M_{\chi}}v_idV\,,
\\[5mm]
\di \Delta \phi = 4\pi G\rho_{\chi}\,,
\end{array}
\label{e1} \eeq where $P_{ik}$ is the pressure tensor for neutralino gas. It is
absent in consequence of second assumption. Last terms in balance equations describe
the momentary going away of annihilation products from the collapsing clump.
Taken into account that annihilation and scattering are cross-channels.
In the local form we have the simple system:
\beq \di
\frac{dM}{dt}= -\frac{3(\si v)_{ann}}{2\pi M_{\chi}}\cdot
\frac{M^2}{R^3}, \qquad  \left(\frac{dR}{dt}\right)^2
=\frac{2GM}{R}\left(1-\beta^2 \frac{R}{R_0}\right), \label{e5}
\eeq
which integrates elementary. Here $R$ is a clump radius; $M=4\pi
\rho_{\chi}R^3/3$ is a clump mass; $R_0=const$ is an initial radius;
$\beta^2 \leq 1$ is the constant which fixed by initial velocity of the
surface clump compression. The results are in the parametric form
\beq
\begin{array}{c}
\di M(\zeta)= M_{(*)}\left[1+\frac{(\si
v)_{ann}}{M_{\chi}}\left(\frac{3\rho_{\chi(*)}}{2\pi
G}\right)^{1/2}\left(\frac{\zeta^3}{3}+\beta^2
\zeta\right)\right]^{-2},
\\[6mm]
\di R(\zeta)=R_0(\zeta^2+\beta^2)^{-1},
\\[6mm]
\di \left(\frac{8\pi
G\rho_{\chi(*)}}{3}\right)^{1/2}t(\zeta)=\frac{1}{\beta^2}\left(\frac{\zeta}{\zeta^2+\beta^2}
-(1-\beta^2)^{1/2}\right)
+\frac{1}{\beta^3}\left(\arctan\frac{\zeta}{\beta}
-\arctan\frac{(1-\beta^2)^{1/2}}{\beta}\right)+
\\[6mm]
\di +\frac{(\si
v)_{ann}}{3M_{\chi}}\left(\frac{3\rho_{\chi(*)}}{2\pi
G}\right)^{1/2}\left(\ln(\zeta^2+\beta^2)
+\frac{2\beta^2(\zeta^2+\beta^2-1)}{\zeta^2+\beta^2}\right),
\end{array}
\label{e6} \eeq where $M_{(*)}\sim M_\odot =4\cdot
10^{33} g,\;\rho_{\chi(*)}\sim 10^{-4} g/cm^3$ are initial clump mass and density. As
you see $M(t)\to
0,\; R(t)\to 0$ for $t \to \infty$.

\begin{figure}[!h]
\begin{minipage}{0.5\textwidth}
\epsfxsize=\textwidth\epsfbox{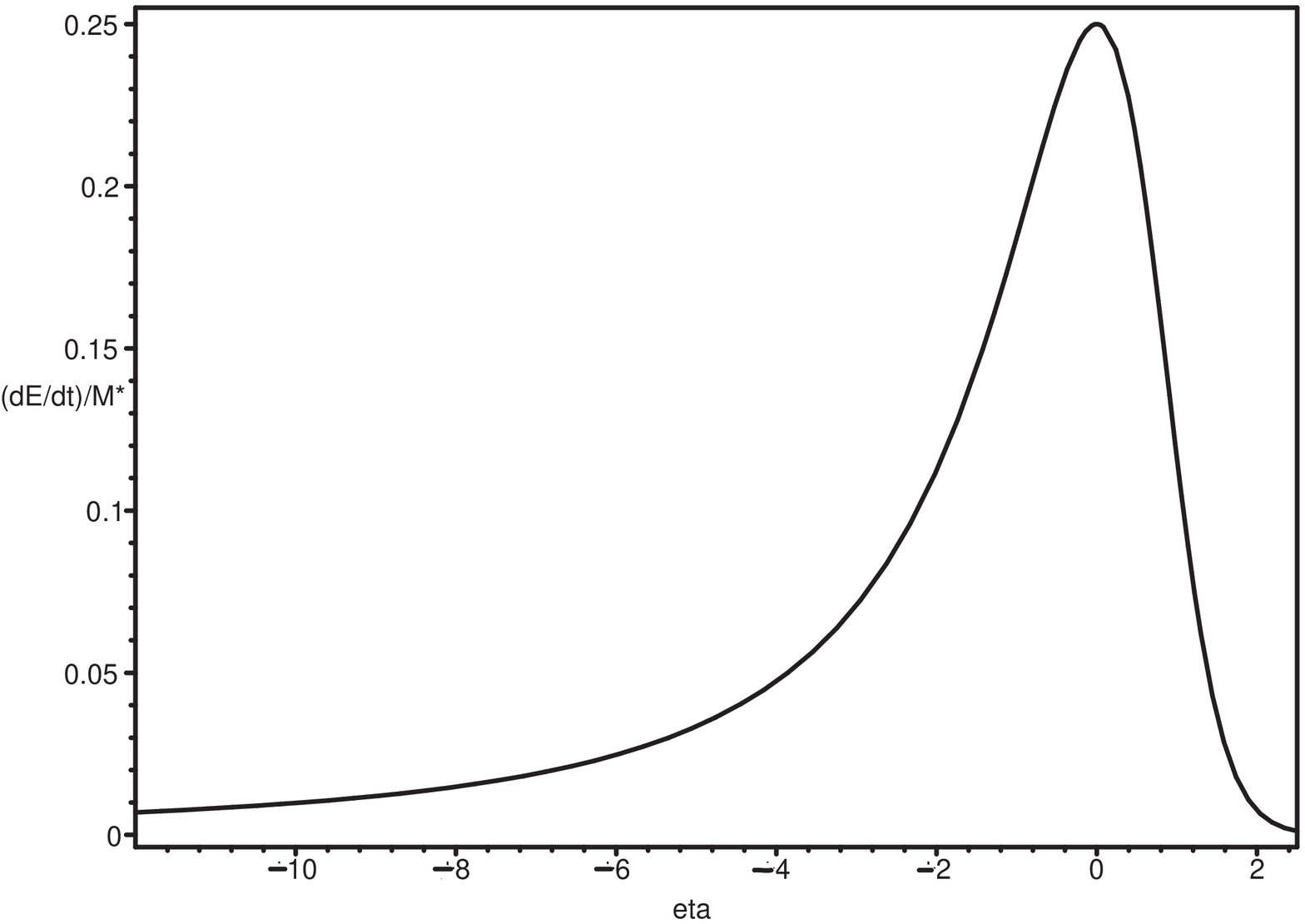}
\end{minipage}
\begin{minipage}{0.5\textwidth}
\epsfxsize=\textwidth \epsfbox{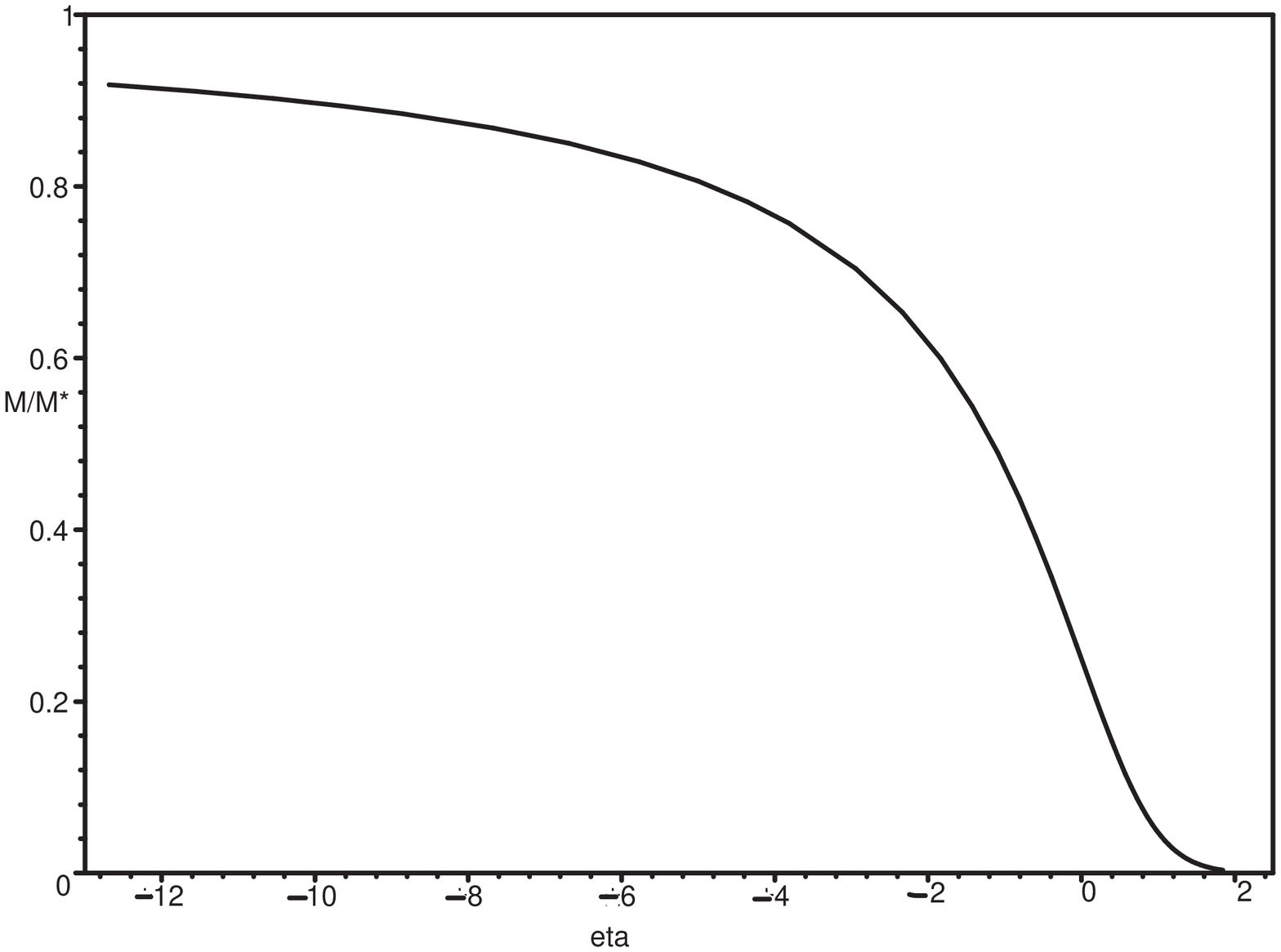}
\end{minipage}
\caption{\small Annihilation rate $\dot E_A/M_{(*)}$ and
relative mass $M/M_{(*)}$ of neutralino clump as a function of
dimensionless time $\eta$.} \label{figH}
\end{figure}

Annihilation rate is \beq \di -\frac{dM}{dt}\equiv \dot{E}_{ann}
=M_{(*)}\rho_{\chi(*)}\cdot \frac{2(\si v)_{ann}}{M_{\chi}} \cdot
\frac{(\zeta^2+\beta^2)^3}{\di \left[1+\frac{(\si
v)_{ann}}{M_{\chi}}\left(\frac{3\rho_{\chi(*)}}{2\pi
G}\right)^{1/2}\left(\frac{\zeta^3}{3}+\beta^2
\zeta\right)\right]^{4}}. \label{rad} \eeq Generally behavior of
this function is that a slow growth at small times changes into
quick evolution around narrow peak after which there is an
exponential drop to zero. This behavior explains by changing of
the parameter hierarchy during the collapse. We use the convenient
time scale \beq \di
\dot{E}_{ann}=\frac{M_{(*)}}{4}\left(1-\frac{(t-t_0)^2}{2\tau_0^2}\right),
\qquad \tau_0= \frac{(\si v)_{ann}}{3\pi G M_{\chi}}. \label{max}
\eeq Here $t_0$ is the time of maximum. As we have
\[
\di g=\frac{(\si
v)_{ann}}{M_{\chi}}\left(\frac{3\rho_{\chi(*)}}{2\pi
G}\right)^{1/2}\ll 1
\]
than the annihilation rate doesn't depend on initial conditions and
expresses through fundamental constants only. Dependence of function $\dot{E}_{ann}/M_{(*)}$
via dimensionless time $\eta = (t-t_0)/\tau_0$ shown at the left Fig. \ref{figH}.
Special features of this function is skewness. About 80 \%
of clump mass annihilates during the time \beq \di t_{ann}=6\tau_0=
\frac{2(\si v)_{ann}}{\pi G
M_{\chi}}=\frac{g_2^4\,(21-10\cos^2{\theta_W}+34\cos^4{\theta_W})}{128\;
\pi^2 M_{\chi}^3 G \cos^4{\theta_W}}\simeq 30 \;s,\qquad
M_{\chi}\simeq 3\;TeV \label{tau} \eeq at that effectively it proceeds between $t_{\scA}^{(-)}=4\tau_0$
(before maximum) and $t_{\scA}^{(+)}=2\tau_0$ (after maximum).

The mass evolution $M_{\chi}$ is shown at the right Fig. \ref{figH}.
Decay time of collapsing neutralino clump expresses through MSSM
parameters and gravitational constant but it is absolutely independent from
initial density and initial velocity of clump compressing.

\section{Conclusion.}

Here we discuss the variant of supersymmetric extension of the SM, where
Higgsino mass scale is lower than the scale of soft SUSY breaking: $\mu\ll
M_{SUSY}\sim M_{0}\sim M_{1/2}$. We call the model as Model with Separated
Higgsinos.

In the model, as in the commonly used MSSM, the precise
convergence of invariant charges takes place at the scale that
does not contradict to proton mean life time constraint. When
$M_{SUSY}>10^7$ GeV radiation corrections dominate in the
neutralino-chargino mass splitting. From the calculation of
splitting and the renormgroup evolution we get
$M_{\chi_1^{\pm}}-M_{\chi_{1,2}^{0}}\simeq 360$ MeV at
$M_{\chi}\simeq \mu \approx 3$ TeV. This value have got from the
analysis of the neutralino relic in cosmological plasma. We
supposed that neutralino is a DM carrier. In the model formation
of the neutralino relic occurs in high symmetric phase of the
plasma. Charged Higgsino mean life time is
$\tau_{\chi_{1}^{\pm}}\sim 0.4\times 10^{-9}$ s. Problems of relic
separated neutralino search in underground (NUSEL) and satellite
(GLAST) experiments are discussed. It have been shown that spin
dependent component dominates in neutralino-nucleon cross section.
Possibility of the separated neutralino registration at
perspective detector XENON is analyzed. In this case the signal is
closed to the planned background. Process of high energy
neutralino recharging on nuclei have been considered for the case
when neutralinos with large kinetic energies (more than 1 TeV) are
in cosmic rays. The spectrum of diffuse annihilation radiation
have been calculated and analyzed for the galactic DM.
Annihilation cross sections for galactic neutralino have been
calculated. Contribution to the gamma ray spectrum from t- and
s-channel neutralino annihilation into $Z^0Z^0$ and $W^{+}W^{-}$
pairs has been calculated using experimental data about hadron
multiplicities at mass surfaces of $Z^0,\,W^{\pm}$-bosons;
quark-antiquark s-channel contribution has been estimated using
the phenomenological model of hadron multiplicities at
$\sqrt{s}=2M_{\chi}.$ In the Galaxy a collapse and total
annihilation of neutralino clumps should take place if neutralinos
are main carriers of its mass. Time and energy distribution of the
gamma ray radiation are calculated. It is shown that nearly 80
percents of the clump mass annihilate in 30 seconds.

\end{document}